\documentclass[prl,twocolumn]{revtex4}
\usepackage{graphicx, epsfig}
\usepackage{color}
\usepackage{mathrsfs}

\newcommand{\be}{\begin{equation}}
\newcommand{\ee}{\end{equation}}
\newcommand{\bea}{\begin{eqnarray}}
\newcommand{\eea}{\end{eqnarray}}

\newcommand{\gapp}{\mathrel{\raise.3ex\hbox{$>$}\mkern-14mu
              \lower0.6ex\hbox{$\sim$}}}
\newcommand{\lapp}{\mathrel{\raise.3ex\hbox{$<$}\mkern-14mu
              \lower0.6ex\hbox{$\sim$}}}

\begin{document}
\title{
Quantum Radiation from Quantum Gravitational Collapse
}
\author{Tanmay Vachaspati$^{1,3}$, Dejan Stojkovic$^{2,3}$}
\affiliation{$^1$Institute for Advanced Study, Princeton, NJ 08540\\
$^2$HEPCOS, Department of Physics, SUNY at Buffalo, Buffalo, NY 14260-1500\\
$^3$CERCA, Department of Physics,
Case Western Reserve University, Cleveland, OH~~44106-7079
}

\begin{abstract}
\noindent
We study quantum radiation emitted during the collapse of a
quantized, gravitating, spherical domain wall. The amount of
radiation emitted during collapse now depends on the wavefunction
of the collapsing wall and the background spacetime. If the wavefunction
is initially in the form of a sharp wavepacket, the expectation
value of the particle occupation number is determined as a
function of time and frequency. The results are in good agreement
with our earlier semiclassical analysis and show that the quantum
radiation is non-thermal and evaporation accompanies gravitational
collapse.
\end{abstract}
%\pacs{???}

\maketitle

%\section{Introduction}
%\label{introduction}

Quantum radiation from gravitationally collapsing matter has been
the subject of a lot of interest
(see {\it e.g.} \cite{Hawking:1974sw,BirrellandDavies,Boulware:1975fe,
Gerlach:1976ji,Hajicek:1986hn} for early work and
\cite{frolov,Ashtekar:2005cj,Townsend} for more recent reviews).
Virtually all of the work in this context has been done in the
semiclassical approximation in which the collapsing matter is
described in classical terms and only the radiation field is quantized.
Such results, though very important, leave open the possibility that
quantum collapse could qualitatively change the outcome. This would
be relevant to both the end-point of Hawking radiation as well as to
the dynamics of gravitational collapse close to the point of black
hole formation. However, a full quantum treatment of gravitational
collapse has been lacking so far.

In Ref.~\cite{tvdslmk} a functional Schrodinger formalism approach
to Hawking evaporation problem was developed, and the quantum evaporation
from a classical collapsing background was studied.
This formalism, being inherently quantum mechanical, is
particularly convenient for treating the collapsing gravitational
background in the context of quantum mechanics. In this paper we  address
the problem where the infalling matter and the radiation field are
both treated in quantum theory.

Our approach follows Ref.~\cite{tvdslmk} and
uses the functional Schrodinger equation
\begin{equation}
H \Psi = i \frac{\partial \Psi}{\partial t}
\label{schrodinger}
\end{equation}
where $\Psi$ is the wavefunctional for a collapsing spherical
domain wall and the excitation modes of the radiation field.
We will find the Hamiltonian more explicitly below,
after which we will solve the equation for the wavefunctional.
The wavefunctional is used to find the excitation spectrum of
the radiation field which leads to the flux of quantum radiation
from the collapsing matter.

% DS commented the next paragraph out
%Our analysis goes an important step beyond semiclassical analyses
%where the background spacetime is treated classically. Here the
%background itself is also quantum. However, the analysis does
%not include the backreaction of the radiation on the spacetime
%and collapsing matter.

%\section{Set up}
%\label{setup}

To study a concrete realization of black hole formation we consider a spherical Nambu-Goto domain wall that is collapsing. To include
the possibility of (spherically symmetric) radiation we consider a
massless scalar field, $\Phi$, that is coupled to the gravitational
field but not directly to the domain wall. The action for
the system is
\begin{equation}
S = \int d^4 x \sqrt{-g} \left [ -\frac{\cal R}{16\pi G}
     + \frac{1}{2} (\partial _\mu \Phi )^2 \right ]
     - {\sigma} \int d^3 \xi \sqrt{- \gamma}
\label{action}
\end{equation}
%\begin{eqnarray}
%S &=& \int d4 x \sqrt{-g} \left [ -\frac{1}{16\pi G} {\cal R}
     %+ \frac{1}{2} (\partial _\mu \Phi )2 \right ] \nonumber \\
     %&-& {\sigma} \int d3 \xi \sqrt{- \gamma}
%\label{action}
%\end{eqnarray}
where the first term is the Einstein-Hilbert action for the
gravitational field, the second is the scalar field action,
the third is the domain wall action in terms of the wall
world volume coordinates, $\xi^a$ ($a=0,1,2$), the wall
tension, ${\sigma}$, and the induced world volume metric
\begin{equation}
\gamma_{ab} = g_{\mu\nu} \partial_a X^\mu \partial_b X^\nu
\label{inducedmetric}
\end{equation}
The coordinates $X^\mu (\xi^a )$ describe the location of the wall,
and Roman indices go over internal domain wall world-volume coordinates
$\zeta^a$, while Greek indices go over space-time coordinates.

We will now only consider spherical domain walls and assume
spherical symmetry for the scalar field ($\Phi = \Phi (t,r)$).
Then the wall is described by only the radial degree of freedom,
$R(t)$. Furthermore, the metric is taken to be the solution of
Einstein equations for a spherical domain wall.
The metric is Schwarzschild outside the wall, as follows
from spherical symmetry \cite{Ipser:1983db}
\begin{equation}
ds^2= -(1-\frac{R_S}{r}) dt^2 + (1-\frac{R_S}{r})^{-1} dr^2 +
      r^2 d\Omega^2 \ , \ \ r > R(t)
\label{metricexterior}
\end{equation}
where, $R_S = 2GM$ is the Schwarzschild radius in terms of the mass,
$M$, of the wall, and $d\Omega^2$ is the usual angular line element.
In the interior of the spherical domain wall, the line element
is flat, as expected by Birkhoff's theorem,
\begin{equation}
ds^2= -dT^2 +  dr^2 + r^2 d\theta^2  + r^2 \sin^2\theta d\phi^2  \ ,
\ \ r < R(t)
\label{metricinterior}
\end{equation}
The interior time coordinate, $T$, is related to the observer time
coordinate, $t$, via the proper time, $\tau$, of the domain wall.
\begin{equation}
\frac{dT}{d\tau} =
      \left [ 1 + \left (\frac{dR}{d\tau} \right )^2 \right ]^{1/2}  ,
 \ \ \
\frac{dt}{d\tau} = \frac{1}{B} \left [ B +
         \left ( \frac{dR}{d\tau} \right )^2 \right ]^{1/2}
\label{bigTandtau}
\end{equation}
where
\begin{equation}
B \equiv 1 - \frac{R_S}{R}
\label{BofR}
\end{equation}
The ratio of these equations connects the interior and exterior
time coordinates
\begin{equation}
\frac{dT}{dt} = \frac{(1+R_{\tau}^2)^{1/2} B}{(B + R_{\tau}^2)^{1/2}}
= \left [ B - \frac{(1-B)}{B} {\dot R}^2 \right ]^{1/2}
\label{tT}
\end{equation}
where $R_{\tau} = dR/d\tau$ and ${\dot R} = dR/dt$.
Integrating Eq.~(\ref{tT}) still requires
knowing $R(\tau)$ (or $R(t)$) as a function of $\tau$ (or $t$).

Note that the wall radius, $R(t)$, completely determines the
metric. Hence, when we quantize $R$, both the wall and the
background will be described quantum mechanically, but the wall
and the background will not have independent quantum dynamics.

Regarding our use of the Schwarzschild time coordinate,
it is important to point out that the Schwarzschild line element
has a coordinate singularity at $r=R_S$. However, the restriction
in Eq.~(\ref{metricexterior}) to $r > R(t)$ means that our coordinate
system is well-defined everywhere as long as $R(t) > R_S$. In all
that follows, we will only be considering the case $R(t) > R_S$.
For example, we will calculate the quantum radiation from a
collapsing wall with radius $R(t) > R_S$.

The energy function for the wall is \cite{tvdslmk}
\begin{equation}
H_{\rm wall} = 4\pi \sigma B^{3/2}R^2 \left [
         \frac{1}{\sqrt{B^2-{\dot R}^2}} -
          \frac{2\pi G\sigma R}{\sqrt{B^2- (1-B){\dot R}^2}}
                                  \right ]
\label{energyfunction}
\end{equation}
where overdots denote derivatives with respect to $t$.
This is a first integral of the equations of motion
and is identified with the energy of the gravitating
domain wall \cite{Ipser:1983db}.

The canonical momentum is given by
\begin{equation}
\Pi \approx \frac{4\pi \mu R^2  {\dot R}}
              {\sqrt{B} \sqrt{B^2-{\dot R}^2}}
\end{equation}
where $\mu \equiv \sigma (1-2\pi G\sigma R_S)$.
Then, in the regime $R \sim R_S$, the wall Hamiltonian is
\begin{eqnarray}
H_{\rm wall}
&\approx& \frac{4\pi \mu B^{3/2}R^2}{\sqrt{B^2-{\dot R}^2}} \label{HRdot}\\
  &=& \left [  (B\Pi)^2 + B (4\pi \mu R^2)^2 \right ] ^{1/2} \label{HPi}
\end{eqnarray}
and has the form of the energy of a relativistic
particle, $\sqrt{p^2 + m^2}$, with a position dependent mass.
In the limit $B \sim 0$, the mass term can be neglected -- the
wall is ultra-relativistic -- and hence
\begin{equation}
H_{\rm wall} \approx - B \Pi
\end{equation}
where we have chosen the negative sign appropriate for describing
a collapsing wall.
%Since we are interested in collapsing walls for which $\Pi$ is
%negative, we will use
%\begin{equation}
%H_{\rm wall} = - B \Pi
%\end{equation}

The scalar field, $\Phi$, is decomposed into a complete set
of basis functions denoted by $\{ f_k (r)\}$
\begin{equation}
\Phi = \sum_k  a_k(t) f_k (r)
\label{modes}
\end{equation}
The exact form of the functions $f_k (r)$ will not be important for us.
We will be interested in the wavefunction for the mode coefficients
$\{ a_k \}$.

The Hamiltonian for the scalar field modes is found by
inserting the scalar field mode decomposition and the background
metric into the action
\begin{equation}
S_\Phi = \int d^4x \sqrt{-g} \frac{1}{2} g^{\mu \nu}
                 \partial_\mu \Phi \partial_\nu \Phi
\end{equation}
The Hamiltonian for the scalar field modes takes the form of coupled
simple harmonic oscillators with $R-$dependent mass and couplings due to the non-trivial metric. In the regime $R \sim R_S$, for a normal
mode denoted by $b$, the Hamiltonian is \cite{tvdslmk}
\begin{equation}
H_{\rm b} = \left ( 1- \frac{R_S}{R} \right ) \frac{\pi^2}{2m}
           + \frac{K}{2} b^2
\end{equation}
where $\pi$ is the momentum conjugate to $b$, $m$ and $K$ are
constants whose precise values are not important for us.

Hence the total quantum Hamiltonian for the wall and a normal
mode of the scalar field is
\begin{equation}
H = H_{\rm wall} + H_{\rm b}
  = - B \Pi + B \frac{\pi^2}{2m} + \frac{K}{2} b^2
\label{Htotal}
\end{equation}
where
$\Pi = -i {\partial}/{\partial R}$,
$\pi = -i {\partial}/{\partial b}$.
%\begin{equation}
%\Pi = -i \frac{\partial}{\partial R} \ , \ \ \
%\pi = -i \frac{\partial}{\partial b}
%\end{equation}
The wavefunction is now a function of $b$, $R$ and $t$:
$\Psi = \Psi (b,R,t)$.
The total Hamiltonian (\ref{Htotal}) is Hermitian with
respect to the inner product with measure $1/B$.

The differential form of the Schrodinger equation is
\begin{equation}
+ i B \frac{\partial \Psi}{\partial R} -
\frac{B}{2m}\frac{\partial^2 \Psi}{\partial b^2}
+ \frac{K}{2} b^2 \Psi = i \frac{\partial\Psi}{\partial t}
\label{tdschrod}
\end{equation}
We look for stationary solutions
\begin{equation}
\Psi (b, R, t) = e^{-iEt} \psi (b,R)
\label{ansatz}
\end{equation}
Then the time-independent Schrodinger equation is
\begin{equation}
-\frac{1}{2m} \frac{\partial^2 \psi}{\partial b^2} +
        \frac{m}{2} \omega^2 b^2 \psi - \epsilon  \psi =
        - i \frac{\partial\psi}{\partial R}
\label{tiseq}
\end{equation}
where $\omega^2 = k/(mB)$ and $\epsilon = E/B$.
To solve Eq.~(\ref{tiseq}) we use the ansatz
\begin{equation}
\psi (b,R) = e^{ -i \int^R \epsilon (R') dR' } \phi (b,R)
           = e^{ -i E u } \phi (b,R)
\label{psiphi}
\end{equation}
where
\begin{equation}
u = \int^R \frac{dR'}{B(R')}
  = R + R_S \ln \left | \frac{R}{R_S} - 1 \right |
\label{uandR}
\end{equation}
This leads to the equation for $\phi (b,\eta)$ where
$\eta \equiv R_S - R$
\begin{equation}
-\frac{1}{2m} \frac{\partial^2 \phi}{\partial b^2} +
   \frac{m\omega^2}{2} b^2  \phi = i \frac{\partial\phi}{\partial \eta}
\label{phieq}
\end{equation}
As discussed in Ref.~\cite{Dantas:1990rk}, this has the implicit solution
\begin{equation}
\phi (b,\eta) = e^{i\alpha(\eta)} \left ( \frac{m}{\pi \rho^2} \right )^{1/4}
      \exp \left [ \frac{im}{2}
      \left ( \frac{\rho_\eta}{\rho} + \frac{i}{\rho^2} \right ) b^2
                    \right ]
\label{phisolution}
\end{equation}
where $\rho_\eta$ is the derivative of the function $\rho (\eta)$ with
respect to $\eta$, and the defining equation for $\rho$ is
\begin{equation}
{\rho_{\eta\eta}} + \omega^2 (\eta) \rho = \frac{1}{\rho^3}
\label{rhoeq}
\end{equation}
where
\begin{equation}
\omega^2 (\eta ) = - \frac{kR}{m\eta} \approx - \frac{kR_S}{m\eta}
\label{omegaR}
\end{equation}
The initial conditions for $\rho$ are taken at some large negative
value of $\eta$ ({\it i.e.} large value of $R$) denoted by $\eta_i$,
so that
\begin{equation}
\rho (\eta_i) = \frac{1}{\sqrt{\omega (\eta_i)}} \ , \ \ \
\rho_\eta (\eta_i ) = 0
\end{equation}
The phase $\alpha$ is defined by
\begin{equation}
\alpha (\eta ) = - \frac{1}{2} \int^\eta \frac{d\eta'}{\rho^2 (\eta')}
\end{equation}
Then Eqs.~(\ref{ansatz}) and (\ref{psiphi}) give
\begin{equation}
\Psi (b,R,t) = e^{-iE(u+t)} \phi (b, \eta)
\label{Psiconvenient}
\end{equation}
with $\phi$ given in Eq.~(\ref{phisolution}) and, as above,
$\eta = R_S -R$.

Next we note that $E$ only enters the solution in the first
exponential factor. So we can superpose stationary solutions to
construct wavepackets. For example,
\begin{equation}
\Psi (b,R,t) = \frac{1}{( \pi\sigma^2 )^{1/4}} e^{-(u+t)^2/2\sigma^2}
                \phi(b,\eta)
\label{Psipacket}
\end{equation}
where $\sigma$ is the width of the wavepacket, is a solution to
the time dependent Schrodinger problem in Eq.~(\ref{tdschrod}).
With the chosen initial conditions for $\rho$ at $\eta_i$,
this solution describes a wavepacket for the collapsing
wall and quantum fields in their ground state at $\eta_i$.

Note that we have normalized the wavepacket with unit measure in
the $u$ coordinate since the Schrodinger evolution preserves this
normalization. Also, $u$ can be expressed in terms of $\eta$ through
Eq.~(\ref{uandR}).

The solution (\ref{Psipacket}) describes quantum radiation from
a quantized collapsing shell of matter. The shell is represented
by a wave packet that has a Gaussian fall-off and is moving toward
the horizon $R_S$ which corresponds to $u \rightarrow -\infty$.
(The wave packet is not strictly Gaussian because the function
$\phi$ also depends on $R$ via $\eta$.) Radiation from
the shell, represented by $\phi(b,\eta)$ depends on the position of
the shell. We want to find the main features of this radiation.

It is worth noting that
\begin{equation}
\Psi (b,R,t) = f(u+t) \phi(b,\eta)
\end{equation}
is a solution for any function $f$, provided $\phi (b,\eta)$
satisfies Eq.~(\ref{phieq}). This solution shows that
wavepackets in $u$ do not spread in time at late times.
Since $dR = B du$ from Eq.~(\ref{uandR}) and $B$ gets
smaller as the wall collapses, the wavepacket gets more
sharply peaked in $R$ with time.

%\section{Occupation Number}
%\label{occupationnumber}

In the semiclassical analysis where $R$ is treated classically,
the wavefunction for the mode amplitudes is decomposed into
suitably chosen basis wavefunctions (discussed below). If a
complete basis is denoted by $\{\phi_n\}$, then the expectation
of the occupation number is
\begin{equation} \label{N}
N = \sum_n n | \langle \phi_n | \Psi \rangle | ^2
\end{equation}

Similar considerations hold for the present case where we are
treating $R$ in quantum theory. The only difference is that the
value of $N$ depends on what value $R$ takes. Hence $N$ is given
by an $R-$ and $t-$ dependent probability distribution, and to
find the expectation value of $N$ at any time, we also need to
integrate over $R$. Therefore,
\begin{equation}
\langle N(t) \rangle = \int du ~ \langle N(R,t) \rangle
               = \int du \sum_n n  | \langle \phi_n | \Psi \rangle | ^2
\label{Nmaster}
\end{equation}
Here we have been careful to integrate over $du$ with unit measure
which is equivalent to integrating over $R$ but with measure $1/B$.
%This follows from the Schrodinger equation, Eq.~(\ref{tdschrod}).

To proceed further, we need to specify our basis wavefunctions, $\phi_n$.
These are chosen to be simple harmonic oscillator basis states at a
frequency ${\bar \omega}$
\begin{equation}
\phi_n (b) = \left ( \frac{m{\bar \omega}}{\pi} \right )^{1/4}
        \frac{e^{-m{\bar \omega} b^2/2}}{\sqrt{2^n n!}}
          H_n (\sqrt{m{\bar \omega}} b)
\end{equation}
where $H_n$ are Hermite polynomials.
Then \cite{tvdslmk}
\begin{equation}
\langle \phi_n | \Psi \rangle =  f(u+t)
 \frac{(-1)^{n/2} e^{-i\alpha}}{({\bar \omega} \rho^2)^{1/4}}
      \sqrt{\frac{2}{P}}  \left ( 1- \frac{2}{P} \right )^{n/2}
      \frac{(n-1)!!}{\sqrt{n!}}
\label{cnresult}
\end{equation}
where $f(u+t)$ can be chosen to be a Gaussian function as in
Eq.~(\ref{Psipacket}) and
\begin{equation}
P \equiv 1 - \frac{i}{\bar \omega}
       \left ( \frac{\rho_\eta}{\rho} + \frac{i}{\rho^2} \right )
\end{equation}

The sum over $n$ in Eq.~(\ref{Nmaster}) can be done explicitly
\cite{tvdslmk} and the occupation number is
\begin{equation}
\langle N(t) \rangle = \int du ~ |f(u+t)|^2 ~
          \frac{{\bar \omega} \rho^2}{\sqrt{2}} \left [
      \left ( 1- \frac{1}{{\bar \omega} \rho^2} \right )^2
     + \left ( \frac{\rho_\eta}{{\bar \omega} \rho} \right )^2
                \right ]
\label{Neq}
\end{equation}
Note that the integrand contains $\rho$ which is a function of
$\eta = R_S-R$ (Eq.~(\ref{rhoeq})), which in turn is a function
of $u$ (Eq.~(\ref{uandR})).
The frequency $\omega$ in Eq.~(\ref{omegaR}) depends on $R$ and
hence is a quantum variable. The frequency ${\bar \omega}$ at which
$N$ is evaluated is chosen to be the expectation value of
$\omega$: ${\bar \omega} = \langle \omega \rangle$.

To evaluate $\langle N \rangle$ we now need to solve Eq.~(\ref{rhoeq})
and then insert the solution in the integrand in Eq.~(\ref{Neq})
together with the Gaussian form of $f(u+t)$.
Finally, we need to do the $u$ integral in Eq.~(\ref{Neq}).
If $f(u+t)$ is sharply peaked, only a small window of $u$,
and an even smaller window of $R$ (Eq.~(\ref{uandR})), contributes
to the integral.

\begin{figure}
\scalebox{0.750}{\includegraphics{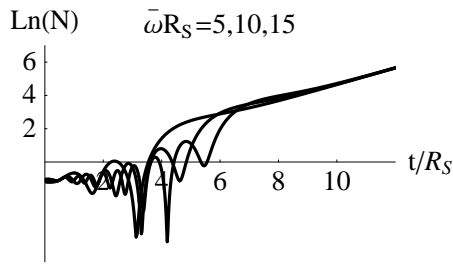}}
%\scalebox{0.750}{\includegraphics{Nvst.eps}}
\caption{$\ln(N)$ versus $t/R_S$ for various fixed values of
${\bar \omega}  R_S$ and for the wavepacket width
$\sigma = 0.1R_S$. Here, ${\bar \omega}$ is the expectation value of
the frequency calculated for the state (\ref{Psipacket}). The linear
growth shows that the occupation number increases exponentially at late
times.
}
\label{Nvst}
\end{figure}

\begin{figure}
\scalebox{0.6}{\includegraphics{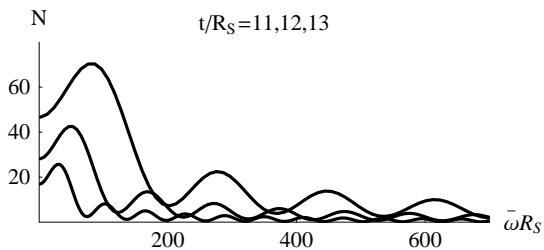}}
%\scalebox{0.7}{\includegraphics{Nvsomega.eps}}
\caption{$N$ versus $ {\bar \omega}  R_S$ for
various fixed values of $t/R_S$ for $\sigma =0.1R_S$. The
oscillations depend on the choice
of width of the wavepacket $\sigma$.  Here, ${\bar \omega}$ is the expectation
value of the frequency calculated for the state (\ref{Psipacket}). The
occupation number at
any frequency grows as $t/R_S$ increases.
}
\label{Nvsomega}
\end{figure}

We have calculated the occupation number numerically for the case
of a wavepacket with $\sigma = 0.1R_S$ and we show the
plot of $\langle N(t) \rangle$ versus $t$ in Fig.~\ref{Nvst}.
These calculations, when done for a range of ${\bar \omega}$,
also yield the spectrum of occupation numbers. In Fig.~\ref{Nvsomega}
we show the spectrum for a few different times.
This result has the same general features as in the semiclassical
analysis \cite{tvdslmk}. The main effect of the quantized background
is that the spectrum becomes more non-thermal though this depends
on the width of the wavepacket. In the case of a very
sharp wavepacket ($\sigma \rightarrow 0$),
the spectrum becomes identical to the semiclassical spectrum.

%\section{Conclusions}
%\label{conclusions}

In summary, we have studied quantum radiation from the changing
spacetime metric due to a collapsing spherical domain wall. The
main improvement over earlier calculations is that the
collapsing matter {\it i.e.} domain
wall is treated within quantum theory.
The end results for the growth of particle occupation number
and spectrum are very similar to that found in semiclassical
analysis \cite{tvdslmk} and, in fact, become identical to the
semiclassical result as the width of the wavepacket decreases.

Our analysis implies that quantum radiation is emitted continuously
during quantum gravitational collapse, and so evaporation and
collapse are concomitant. As seen in Fig.~\ref{Nvst}, the
occupation number at any frequency grows with time, and as
in Fig.~\ref{Nvsomega}, the spectrum has non-thermal features.
In an earlier paper \cite{tvdslmk}, we have approximately fitted
a thermal distribution to the spectrum in an intermediate range
of frequencies, and found a radiation roughly consistent with the
Hawking temperature.
%$T \sim 2.5 T_H$ where $T_H = 1/(4\pi R_S)$
%is the Hawking temperature.
As the collapse proceeds, the window of frequencies in which the
radiation is thermal grows, but only in the $t \rightarrow \infty$
limit, does the spectrum become thermal in an infinite range
of frequencies. The thermal spectrum, however, may never be
achieved in a physical setting precisely because it is only
realized in the $t \rightarrow \infty$ limit.
It is also very important that small non-thermalities
found in the semi-classical treatment \cite{tvdslmk} get
amplified in the full quantum treatment.

We should note that our analysis has ignored radiation backreaction
on the collapse process since we take $R_S$ to be constant. A full
study of backreaction must involve time dependence of $R_S$.

\begin{acknowledgments}
This work was supported by the U.S. Department of Energy and
NASA at Case Western Reserve University.
 Work of DS was also supported by HEPCOS group, SUNY at Buffalo.

\end{acknowledgments}

%\appendix

%\section{$\rho$ equation}
%\label{rhoeqdiscussion}

\end{document}